\documentclass[letter, 10pt]{sig-alternate-10pt} 
\pdfoutput=1
\usepackage{tikz}
\usepackage{framed,multicol}
\usetikzlibrary{shapes,arrows,fit,calc,positioning}
\usepackage{cite}
\usepackage{pgfplots}
\usepackage{pgfplotstable}
\usepackage{multirow}
\usepackage{tabularx}
\usepackage{algorithm}
\usepackage{algorithmic}
\usepackage{xspace}
\usepackage{subcaption}
\usepackage{array,graphicx}
\usepackage{booktabs}
\usepackage{pifont}
\usepackage{balance}
\usepackage{epsfig}
\usepackage{amsmath}
\usepackage{mathtools}
\usepackage{url}
\usepackage{mathptmx}
\usepackage{balance}
\usepackage{scalefnt}

\usetikzlibrary{pgfplots.groupplots}
\usetikzlibrary{patterns}

\definecolor{colorbrewer1}{RGB}{213,62,79}
\definecolor{colorbrewer2}{RGB}{252,141,89}
\definecolor{colorbrewer3}{RGB}{254,224,139}
\definecolor{colorbrewer4}{RGB}{230,245,152}
\definecolor{colorbrewer5}{RGB}{255,153,51}
\definecolor{colorbrewer6}{RGB}{0,82,204}
\definecolor{colorbrewer7}{RGB}{240,248,255}

\newcommand{\tabref}[1]{{Table~\ref{#1}}}
\newcommand{\figref}[1]{{Figure~\ref{#1}}}

\newcommand{\bheading}[1]{{\vspace{4pt}\noindent\textbf{#1}}}

\makeatletter
\renewcommand*{\thetable}{\arabic{table}}
\renewcommand*{\thefigure}{\arabic{figure}}
\let\c@table\c@figure
\makeatother 

\newlength{\saveparindent}
\newlength{\saveparskip}
\newcounter{ctr}

\newenvironment{newitemize}{%
\begin{list}{\mbox{}\hspace{5pt}$\bullet$\hfill}{\labelwidth=15pt%
\labelsep=5pt \leftmargin=20pt \topsep=3pt%
\setlength{\listparindent}{\saveparindent}%
\setlength{\parsep}{\saveparskip}%
\setlength{\itemsep}{3pt} }}{\end{list}}

\newlength\figureheight
\newlength\figurewidth
\newlength{\arrow}
\newlength{\arrowl}
\makeatletter
\newcommand{\mydashleftrightarrow}[2][]{\ext@arrow 3359\leftrightarrowfill@@{#1}{#2}}
\def\rightarrowfill@@{\arrowfill@@\relax\relbar\rightarrow}
\def\leftarrowfill@@{\arrowfill@@\leftarrow\relbar\relax}
\def\leftrightarrowfill@@{\arrowfill@@\leftarrow\relbar\rightarrow}
\def\arrowfill@@#1#2#3#4{%
  $\m@th\thickmuskip0mu\medmuskip\thickmuskip\thinmuskip\thickmuskip
   \relax#4#1
   \xleaders\hbox{$#4#2$}\hfill
   #3$%
}
\makeatother

\newcommand{\clnode}{$C$\xspace}

\newcommand{\code}[1]{\texttt{#1}}
\newcommand{\paragraphq}[1]{\noindent\textbf{#1.}}

\def\showcomments{1}
\ifnum\showcomments=1
    \newcommand{\fixme}[1]{{\textcolor{red}{[FIXME: #1]}}}
    \newcommand{\checkme}[1]{{\textcolor{orange}{[CHECKME: #1]}}}
    \newcommand{\liangw}[1]{{\textcolor{red}{[LW: #1]}}}
    
    \newcommand{\tnote}[1]{{\textcolor{blue}{[TomR: #1]}}}
    \newcounter{mynote}[section]
    
    \newcommand{\thenote}{\thesection.\arabic{mynote}}
\else
    \newcommand{\fixme}[1]{}
    \newcommand{\checkme}[1]{}
    \newcommand{\thenote}[1]{}
    \newcommand{\liangw}[1]{}
    \newcommand{\tnote}[1]{}
\fi

\title{\LARGE \bf Towards Better Understanding of Bitcoin Unreachable Peers}

\author{Liang Wang, Ivan Pustogarov\\ UW-Madison, Cornell Tech\\liangw@cs.wisc.edu, ivan.pustogarov@cornell.edu}

\begin{document}

\maketitle
\thispagestyle{empty}
\pagestyle{empty}

%%%%%%%%%%%%%%%%%%%%%%%%%%%%%%%%%%%%%%%%%%%%%%%%%%%%%%%%%%%%%%%%%%%%%%%%%%%%%%%%
\begin{abstract}
The bitcoin peer-to-peer network has drawn significant attention from
researchers, but so far has mostly focused on publicly visible portions of the
network, i.e., publicly reachable peers. This mostly ignores the hidden parts
of the network: unreachable Bitcoin peers behind NATs and firewalls. In this
paper, we characterize Bitcoin peers that might be behind NATs or firewalls
from different perspectives. Using a special-purpose measurement tool
we conduct a large scale measurement study of the Bitcoin
network, and discover several previously unreported usage patterns: a small
number of peers are involved in the propagation of 89\% of all bitcoin
transactions, public cloud services are being used for Bitcoin network probing
and crawling, a large amount of transactions are generated from only two mobile
applications.
We also empirically
evaluate a method that uses timing information to
re-identify the peer that created a transaction against unreachable peers. 
We find this method very accurate for peers that use the latest version of
the Bitcoin Core client. 
\end{abstract}

\section{Introduction}

Bitcoin~\cite{nakamoto2012bitcoin} is a cryptocurrency and a peer-to-peer
network. The explosion in its popularity is fueled in large part by its
decentralized nature, low transaction fees, and ease in participating. 
A whole Bitcoin ecosystem, the core of which is the Bitcoin network, 
exists today. Understanding its properties and usage patterns of its 
participants can give insights into improving the Bitcoin network, as well as the ecosystem. 

Known research on the Bitcoin network has mostly focused on the publicly
visible part of the network, i.e., publicly reachable peers, and ignored the
hidden part -- clients behind NATs and firewalls that do not allow inbound 
connections~\cite{decker2013information,bbbb,cccc,dddd,eeee}. Nonetheless, gathering statistics on such peers is equally, if not
even more important, since the number of such peers is estimated to be to order of
magnitude larger than the number of reachable peers. 
In this paper we conduct a large scale measurement to collect different
statistics and analyze the usage patterns of unreachable Bitcoin peers that might 
hide behind NAT or firewalls. To facilitate our measurement, we
designed and implemented an open-source measurement tool that call we \emph{bcclient}, 
which can serve as a Bitcoin node but with extended features to aid in 
collecting specific connection and transaction information. 
We conduct a measurement study using bcclient using 102 Bitcoin nodes, which were distributed over 
14 geographical regions spanning all continents, in the Bitcoin main network 
for seven days, and collected information on about 3\,M connections and 
2.5\,M unique transactions. These connections are generated from about
190\,K IPv4 IPs, and we used bcclient to discover that 87\% of these IPs
can
be associated with unreachable peers. Our further analysis on collected data 
suggests:   

\begin{newitemize}

\item Bitcoin unreachable peers appear to be centralized in terms of 
Internet routing and transaction propagation: 16\% of these peers are hosted 
in only 5 Autonomous Systems (which constitues only 0.07\% of all observed
ASes), and 50 unreachable peers are involved in propagation of 
43\% of transactions. 

\item About 80\% of unreachable peers are associated with two mobile Bitcoin 
applications, which contribute to 61\% of unique transactions and 20\% of 
connections. 

\item A large fraction of connections came from IPs in public clouds; %which contains 
a considerable amount of these connections might be used for Bitcoin network 
probing and crawling.

\end{newitemize}

Inspired by our measurement results, we further exercised bcclient and
an experimental framework to empirically evaluate a variation of a known
method that uses timing information to re-identify the peer that created
a transaction against unreachable peers.
This method can be useful to further enrich the collected dataset with
valuable statistics, such as number transaction generated by each
country. 
In our experiments we were able to re-identify every second transaction
generated by us with the latest at the time of writing version of Bitcoin Core
software~(v0.14.1). We observed negligible false positive rates.
We have made the source code for bcclient publicly available to facilitate
future research.

\noindent \textbf{Roadmap:}
We start by providing the necessary background to understand Bitcoin
peer-to-peer network and message propagation rules in
Section~\ref{sec:background}. In Section~\ref{sec:stats}, we present our
custom-built Bitcoin software, describe our large scale experiments, and
present statistics on Bitcoin clients. In Section~\ref{sec:timing},
we empirically evaluate our peer triage method.
Section~\ref{sec:conclusion} concludes the paper.

\section{Background}
\label{sec:background}
%\subsection{Bitcoin P2P network}
\paragraphq{Bitcoin P2P network}
Bitcoin is a digital decentralized currency that relies on cryptography
and a peer-to-peer network for double-spending prevention instead of a
trusted third party. 
%Bitcoin is pseudonymous, creating a Bitcoin address
%does not require any form of identification, and one can create any
%number of addresses locally. 
%: they are merely SHA-256 hashes of user-generated ECDSA public keys.
Bitcoin users can pay each other by creating cryptographically signed
\textit{transactions} and broadcasting them in the Bitcoin peer-to-peer
network. In order to facilitate transaction propagation the network
implements a simple gossip protocol: every peer that received a message
forwards it to its neighbors.

The core of the Bitcoin network is a set of about 7000 servers with
public IP addresses operated by volunteers and companies. The network is
open and anybody can run a Bitcoin server and contribute with  bandwidth
and computational resources. Bitcoin users (often behind firewalls
and/or NAT's) that do not allow inbound connection access the network,
%through these servers, Thus Bitcoin servers are essential for keeping
%the Bitcoin system operational: clients use it to connect to the
%network,
send transactions, and learn about transactions of others through these
servers. The list of servers is publicly available and the default
behavior for a server is to accept up to 117 inbound connections on TCP
port 8333. The default behaviour for a Bitcoin client is to maintain
connections to 8 different servers.
After Bitcoin peers establish a TCP connection, they complete a
handshake protocol by exchanging \code{VERSION} messages, containing
among other fields the software version/name.

%At the time of writing the Bitcoin network consists of about 7,000 peers
%with public IP addresses and a number of  peers behind firewalls and/or
%NAT's that do not allows inbound connections.  The default behavior for
%a Bitcoin peer is to establish 8 outbound connections to other peers in
%the network, and accept up to 117 additional inbound connections. As
%Bitcoin peers behind NAT cannot accept incoming connections, this leaves
%them with only 8 outbound connections.  The default TCP port for the Bitcoin
%protocol is 8333. 
%After Bitcoin peers establish a TCP connection, they
%complete a handshake protocol by exchanging \code{VERSION}
%messages, containing among other fields the software version/name.

\paragraphq{Transaction forwarding}
%A user can initiate money transfer by creating a \textit{transaction}
%and broadcasting it in the Bitcoin peer-to-peer network which
%implements a simple gossip protocol: every peer that received a
%transaction forwards it to its neighbors (in this way the transaction
%is propagated to every node in the network).  A transaction would
%usually include sending and receiving addresses and the amounts to be
%transferred.
A common transaction consists of the sender's and recipient's Bitcoin
addresses, the sender's public key and the signature. This makes the
transaction pseudonymous as there is not any identifying information
besides a randomly looking public key/address of the user who generated it.
One possible way to identify the sender though is to monitor the network
traffic, and to look at the IP address from where the transaction was first
sent. In order to make such traffic analysis harder and improve users'
privacy, the Bitcoin protocol defines special rules when forwarding a
transaction.  The goal of these rules is to make a transaction generated by a user
indistinguishable from transactions generated by others. % users.  

First, according to the Bitcoin reference implementation whenever a
Bitcoin peer (either a client or a server) receives a transaction from
one of its neighbors, it broadcasts it further to the rest of its
neighbors. In this way the peer's own transactions get mixed with the
ones it relays.

Second, when a peer generates or receives a new transaction it does not
relay it immediately to its neighbors. Instead it chooses and assigns a
random exponentially distributed delay for each of its neighbors; the
actual transmissions occur when the corresponding delays expire. This
impedes timing analysis for an attacker.  The mean of exponential
distribution is different for inbound (5 seconds) and outbound (2.5
seconds) connections.  The random exponentially distributed delays were
introduced in the most recent (at the time of writing) version of Bitcoin
reference implementation and previous
research~\cite{koshyanalysis,juhasz} did not take this modification into
account. 
%Besides random delays, Bitcoin peers use a mechanism called
%``trickling'' which might further delay forwarding transactions. A peer
%would round robin over its neighbors with 100 millisecond intervals and
%would forward transactions to a neighbor only if it is its turn and the
%random delay expired. 

\paragraphq{Bitcoin Testnet}
In order to facilitate testing of new features, the Bitcoin community
runs a small separate testing network called \textit{Testnet} with an
independent Blockchain. At the time of submission Testnet consisted of
about 250 nodes with public IP addresses. By analogy we call the actual
Bitcoin network \textit{Mainnet}.

\newpage
\section{Measurements}
\label{sec:stats}
\subsection{Methodology and dataset}
Most of the existing research on the Bitcoin P2P network focuses on its
backbone: the servers with public IP addresses. Such research is
facilitated by the fact that one can readily connect to the servers and
start collecting the data. In this paper we go one step deeper and try
to collect data on the less visible (but nonetheless large) part of the
Bitcoin ecosystem: clients behind NATs and firewalls that do not accept
incoming connections. We refer to them as \textit{unreachable peers}.
During our measurement we injected more than 100
Bitcoin servers with public IPs distributed over the globe into the
network and were collecting data on incoming connections. We were mostly
interested in the following:
\begin{itemize}
  \item The number of incoming connections over time and number of
	unique IP addresses. We use these figures to estimate the
	number of Bitcoin clients behind NATs and firewalls. 
  \item Location and Autonomous Systems of the discovered IP addresses.
	This information can be used to better understand
	which countries use Bitcoin the most and whether Bitcoin clients
	are centralized in terms of Internet routing.
  \item Type of software used by Bitcoin clients. This can give us some
        insights on whether Bitcoin clients use smartphones to access
        the network, and whether clients access Bitcoin through Tor.
  \item Volume and properties of transactions generated by clients which
        can be used to infer Bitcoin clients usage patterns.
\end{itemize}

\bheading{Bcclient.}
In order to facilitate our measurements we developed a special-purpose
measurement tool, called \emph{bcclient}, using libbitcoin
library~\cite{libbitcoin}.  In a nutshell, 
bcclient is a customized, lightweight Bitcoin software that can either 
run as Bitcoin full node or client, while offering extended functionality such 
as (1) recording specified events of users' interest with precise timestamp, 
(2) establishing several parallel connections to a given Bitcoin node, (3) running 
as a full node without downloading any part of Blockchain, and more. 
Bcclient was optimized for memory and storage usage, and for processing huge 
volumes of data, which make it suitable for conducting large-scale and 
long-term measurement experiments in the main Bitcoin network. Bcclient also 
includes an analysis engine that contains a set of tools for analyzing data. 
The code of bcclient was released publicly to 
facilitate future research\footnote{https://github.com/ivanpustogarov/bcclient}. 

\begin{figure}[t]
\centering
\includegraphics[scale=0.3]{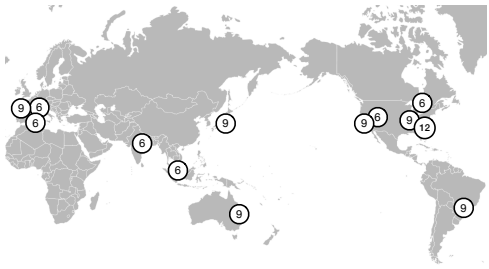}
\caption{Number of measurement nodes in different locations.}\label{fig:deploy}
\end{figure}

\bheading{Data collection.}
For our data collection we set up a total of 102 AWS EC2 servers
running in diverse networks around the world, as shown in
\figref{fig:deploy}~\cite{ec2prex22}.  More specifically, we set up
servers in each of the availability zones in each AWS EC2
region~(that correspond to different geographical locations). 
On each of the EC2 servers we were running an instance of bcclient. We
were logging detailed information on inbound connections from IPv4 addresses
such as connection IP, timing, transaction information, etc. 
We  periodically collected logs from all bcclient instances, and
extracted and deduplicated source  IP addresses of inbound connections.  

During our study we were interested primarily in clients behind NATs and
firewalls, and thus for each extracted IP address we checked if we were
able to establish a reverse connection back to the client.  We sent a
TCP SYN probe to its port 8333~(Bitcoin default), tried to complete the
Bitcoin handshake protocol if the port was open, and recorded the
connection result, i,e., success or fail. 

We have run the measurement three times in March and May 2017. In this paper, 
we use the 60\,G of data collected from the latest 168-hour measurement 
from May 10, 2017 to May 17, 2017. 

\noindent\textit{Ethical considerations.} To avoid adding burden on the
peers being scanned, we only tried to establish a reverse connection to
an IP once in every 6 hours, and would close a connection immediately
after the connection was established successfully. We used a special version
string in bcclient so operators of Bitcoin public peers could find and
contact us if they did not want to be probed.

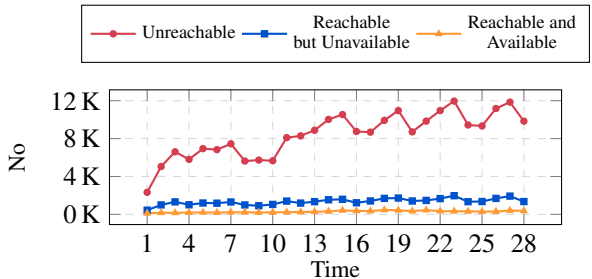
\begin{figure}[t]
\centering
\begin{tikzpicture}
\begin{axis}
[
%legend columns=3, 
%%%basic
yticklabel style={
        /pgf/number format/fixed,
        /pgf/number format/precision=5
},
scaled y ticks=false,
yticklabel = {
    \pgfmathparse{\tick/1000}
    \pgfmathprintnumber{\pgfmathresult}\,K
},
legend style={cells={align=center},at={(0.5,1.5)},anchor=south},
width = 0.9\linewidth, 
height = .4\linewidth,
ylabel = No,
xlabel = Time,
xlabel style={name=xlabel, yshift=0.5em},
ylabel style={name=ylabel, yshift=0em},
%%%set up grid
xtick={1,4,...,28},
ytick={0,4000,...,12000},
label style={font=\footnotesize},
grid = major,
grid style={dashed, gray!30},
legend style={font=\scriptsize, at={(0.5,1.2)}},
legend columns=-1,
]
\addplot+[mark options={mark size=1pt}, line width=0.8pt, mark=*, color=colorbrewer1]  
  table[x=tm, y=ip0, col sep=comma] {figure/ip_conn_tx_time.csv}; 
  \addplot+[mark options={mark size=1pt},line width=0.8pt, mark=square*, color=colorbrewer6] 
  table[x=tm, y=ip1, col sep=comma] {figure/ip_conn_tx_time.csv}; 
  \addplot+[mark options={mark size=1pt},line width=0.8pt, mark=triangle*, color=colorbrewer5] 
  table[x=tm, y=ip2, col sep=comma] {figure/ip_conn_tx_time.csv}; 
\legend{Unreachable, Reachable \\ but Unavailable , Reachable and \\ Available}

\end{axis}
\end{tikzpicture}

% \begin{tikzpicture}
% \begin{axis}
% [
% %%%basic
% % yticklabel style={
% %         /pgf/number format/fixed,
% %         /pgf/number format/precision=1
% % },
% scaled y ticks=false,
% % yticklabel = {
% %     \pgfmathparse{\tick/100000}
% %     \pgfmathprintnumber{\pgfmathresult}\,K
% % },
% legend style={at={(0.5,1.5)},anchor=north},
% width = 0.9\linewidth, 
% height = .4\linewidth,
% ylabel = No,
% xlabel = Time,
% xlabel style={name=xlabel, yshift=0.5em},
% ylabel style={name=ylabel, yshift=0em},
% label style={font=\footnotesize},
% xtick={1,4,...,28},
% ytick={5000,20000,40000,60000},
% ymin=5000,
% ymax=60000,
% grid = major,
% grid style={dashed, gray!30},
% % legend style={font=\footnotesize},
% % legend pos=south east,
% legend columns=-1,
% ]
% \addplot+[mark options={mark size=0.5pt},line width=0.8pt] 
%   table[x=tm, y=conn0, col sep=comma] {figure/ip_conn_tx_time.csv};
%   \addplot+[mark options={mark size=0.5pt},line width=0.8pt] 
%   table[x=tm, y=conn1, col sep=comma] {figure/ip_conn_tx_time.csv};
%   \addplot+[mark options={mark size=0.5pt},line width=0.8pt] 
%   table[x=tm, y=conn2, col sep=comma] {figure/ip_conn_tx_time.csv};
% \legend{Unresponisve, Unavailable, Available}
% \end{axis}
% \end{tikzpicture}
\caption{Change over time in different types of IPs that our nodes have seen. 
Each point on X-axis represents a 6-hour period.}\label{fig:ip_conn_tx_time}
\end{figure}

\bheading{Overview of dataset.}
During our experiments we observed a total of 2,956,515 inbound
connections from 189,204 unique source IP addresses;
for 99.6\% of these connections, we had the timing
information: both their opening times and closing times in our logs. We
say such connections are \emph{completed}

Our measurement nodes have received a total of 2,490,042 unique
transactions from 4,516 IPs. A transaction might be relayed to our nodes
multiple times by different peers; we refer to each transaction receive
event as \emph{propagation}. We saw 83,878,269 propagations.

\subsection{Characterizing Bitcoin clients}
\bheading{Reachable vs. unreachable.}
We observed that most of the incoming connection come from peers that do
not accept incoming connections (i.e. unreachable peers).

We extracted a total of 189,204 unique source IP addresses (IPv4) from
the completed connections and categorized them into groups based
on whether they responded to our probes. 

\noindent\emph{Unreachable IPs}.
We say an IP address is unreachable if it fails to respond to any of our
TCP probes during the measurement.  We found that 86.8\% of the
collected IPs are unreachable.  We mainly focus on unreachable IPs in
our analysis, but also report on statistics about other types of IPs for
comparison purposes in some cases.

\noindent\emph{Reachable IPs}. We say that an IP address is reachable if we
were able to successfully establish a TCP connection at port 8333 back
to such IP address. We found only 13.2\% of IPs to be reachable under this
definition. 
We further split reachable peers into those that successfully completed the
Bitcoin handshake protocol (we call them \emph{available}) and those that did not
(we call them \emph{unavailable}).  We found only 1,587~(0.8\%) IPs
successfully completed the handshake. We further checked them
against the IPs of public Bitcoin nodes provided by~\cite{bitnodes}, and
found that 1,073 of available IPs belong to the set of public Bitcoin servers.
Table~\ref{dataset} further shows the breakdown of number of
connections, transactions, and propagations from different type of IP
addresses. 

\noindent\emph{Remark.}
It is possible for peers to use a non-default port in which case we would mark
them as unreachable.

\bheading{Estimating number of clients.}
\figref{fig:ip_conn_tx_time} shows how the number of reachable and
unreachable IPs changed over time. At the beginning, the number of
unreachable peers gradually increases; this is due to the fact our
Bitcoin servers were new to the network and Bitcoin clients were learning their
IP addresses over time. After 78 hours the number of unreachable peers
become relatively stable with the average about 10K peers over 6 hours
periods. This can be extrapolated to estimate the total number of
Bitcoin peers behind NATs and firewalls. Given that an unreachable
client establishes 3.5 parallel connections to the network on the average (see
below) and that there were 5,540 Bitcoin servers with IPv4 address on
the average during our experiment, we estimate the number of unreachable
clients to be at least 155,000 at any given 6-hours time interval. 

\noindent\emph{Remark.}
Note that this number assumes that there is one-to-one correspondence between
an IP address and a client.  Different clients however can share the same NAT
address, and the real number of unreachable clients might be bigger.  An
unreachable or unavailable IP could be used as a \emph{frontend} IP, which is
for the gateway of a network, or the frontend of a cluster of peers.

\bheading{Client version.}
We extracted a total of 534 unique client version strings from collected
connections, and found that the most popular three Bitcoin client
software across all IPs are BitcoinWallet, Breadwallet, and Bitcore,
which are associated with 59.9\%, 20.0\%, and 6.9\% of all IPs
respectively.  Unreachable and unavailable IPs are mostly associated
with version strings ``Bitcoin Wallet'' or ``breadwallet''. We looked at
the their source code and found that BitcoinWallet establishes 3
parallel connections to the network; Breadwallet establishes either 4 or
6 parallel connections depending on the amount of RAM on the device.

Available IPs are mostly associated with version strings ``Satoshi'' or
``BitcoinUnlimited''. We also found 17,144 unreachable IPs (19,813 of
all IPs) have sent empty version strings. Note that we saw 13\% of IPs
were associated with multiple version strings which suggests that these
are different users behind the same NAT address. 

We found many users with outdated versions of these apps: only 48.3\% of
BitcoinWallet and 79.4\% of Breadwallet clients are using the latest
releases. Surprisingly, 98.9\% of Bitcore clients uses outdated versions
of client software. 
% 113300, 37786, 13086

\begin{table}[t]
\footnotesize
\centering
\begin{tabular}{|r|r|r|}
\hline
 \textbf{Type} & \multicolumn{1}{c|}{\textbf{\# IP}} & \multicolumn{1}{c|}{\textbf{\# Conn}} \\ \hline
 \hline 
0     & 164,198~(86.8) & 936,845~(31.8)   \\ \hline
1     & 23,419~(12.4)  & 1,095,445~(37.2) \\ \hline
2     & 1,587~(0.8)    & 911,173~(31.0)   \\ \hline
Total & 189,204~(100)  & 2,943,463~(100)  \\ \hline  
\hline
 \textbf{Type} & \multicolumn{1}{c|}{\textbf{\# Prop}} & \textbf{\# TX} \\ \hline
 \hline 
0     & 40,071,909~(47.8) & 2,409,087\\ \hline
1     & 8,987,300~(10.7)  & 8,987,300\\ \hline
2     & 34,819,060~(41.5) & 2,354,397 \\ \hline
Total & 83,878,269~(100)  & 2,490,042 \\ \hline  
\end{tabular}
%}
\caption{An overview of dataset. ``Conn'',``TX'' and ``Prop'' stand for connection, 
unique transaction ID, and propagation, respectively. Type 0, 1, 2 
correspond to unreachable, unavailable, available IPs, respectively. Number 
in parenthesis is the percentage of the total number in the last row.}
\label{dataset}
\end{table}

\bheading{Centralization of Bitcoin clients.}
We checked geolocation (i.e., country) and Autonomous Systems (ASes) of
all the collected IPs using a tool called pyasn~\cite{pyasn}.  We
observed that Bitcoin peers are highly centralized in terms of Internet
routing~\cite{apostolakihijacking} and transaction propagation.

(1)~\emph{Peers location.} 
The observed connections coming from 7,070 distinct ASes that span all
countries.  The country with most IPs is United States, which has 24.2\%
of collected IPs, followed by Germany~(7.3\%), Canada~(5.5\%),
Russia~(4.9\%), and United Kingdom (4.7\%). 

(2)~\emph{Autonomous systems.} 
We found that just 5 ASes (which constitues only 0.07\% of all observed
ASes) host as much as 16.0\% of the unreachable peers; these top 5 ASes are
associated with T-Mobile, Comcast, Verizon, ATT, and Rogers which
suggests that associated peers might run on mobile devices.  Furthemore
100 ASes (which constitues only 1.4\% of all observed ASes)  host as much as
60.1\% of oberved IPs.

We also analyzed how active each AS was.  
We found that 63.5\% of completed
connections were generated from only 10 ASes.  Furthermore  we found
that near 99\% of the ASes are only associated with a small number of
connections -- each of these ASes generated  less than 0.1\% of
completed connections.  For connections from unreachable IPs the result
is similar: 33\% of them are from 10 ASes. 

(3)~\emph{Transaction propagation.} We observed that
a small number of IPs are associated with most  of the transaction
propagations. More specifically, 100 IPs are associated with 89\% of all the
propagations; and 50 unreachable IPs are involved in 43\% of all the
propagations.  These IPs could become the ``bottleneck'' of Bitcoin networks,
i.e., taking down one of these IPs can affect propagation of a considerably
large number of transactions. Inspecting the top ten IPs in terms of relayed
transactions manually, five of them are public Bitcoin nodes; however, the
other IPs seems not be frontends of services or gateways of networks. This
leaves why they involved in so many transactions an open question and we are in
process of inspecting it.

\bheading{Remark.}
Centralization in Bitcoin peers might cause security and privacy
issues~\cite{apostolakihijacking}.  For example, most of unreachable
peers are associated to two mobile Bitcoin applications, which means
that security of a large number of transaction is in hands of just two
companies.

\subsection{Understanding usage patterns.}

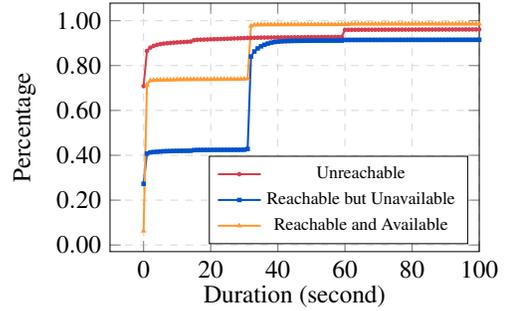
\begin{figure}[t]
\centering
\resizebox{.8\columnwidth}{!}{\begin{tikzpicture}
\begin{axis}
[
y tick label style={
        /pgf/number format/.cd,
            fixed,
            fixed zerofill,
            precision=2,
        /tikz/.cd
    },
%%%basic
width = 0.8\linewidth, 
height = .6\linewidth,
ylabel = Percentage,
xlabel = Duration (second),
xlabel style={name=xlabel, yshift=0.5em},
ylabel style={name=ylabel, yshift=-0.1em},
%%%set up grid
xmax = 100,
grid = major,
xtick={},
ytick={0,0.2,...,1},
grid style={dashed, gray!30},
%legend style={font=\footnotesize},
%legend style={cells={align=center},font=\footnotesize},at={(0.5,1.5)},anchor=south},
legend style={cells={align=center},font=\scriptsize},
legend pos=south east,
]
\addplot+[mark options={mark size=0.5pt},line width=0.8pt, mark=*, color=colorbrewer1] 
  table[x=x, y=y, col sep=comma] {figure/conndur_tp_0.csv};
  \addplot+[mark options={mark size=0.5pt},line width=0.8pt, mark=square*, color=colorbrewer6] 
  table[x=x, y=y, col sep=comma] {figure/conndur_tp_1.csv};
  \addplot+[mark options={mark size=0.5pt},line width=0.8pt, mark=triangle*, color=colorbrewer5] 
  table[x=x, y=y, col sep=comma] {figure/conndur_tp_2.csv};
%\legend{Unresponisve, Unavailable, Available}
%\legend{Unreachable, Reachable but \\ Unavailable, Reachable and \\ Available}
\legend{Unreachable, Reachable but Unavailable, Reachable and Available}
\end{axis}
\end{tikzpicture}

% \begin{tikzpicture}
% \begin{axis}
% [
% %legend columns=3, 
% y tick label style={
%         /pgf/number format/.cd,
%             fixed,
%             fixed zerofill,
%             precision=2,
%         /tikz/.cd
%     },
% %%%basic
% width = 0.8\linewidth, 
% height = .6\linewidth,
% ylabel = Percentage,
% xlabel = Duration,
% xlabel style={name=xlabel, yshift=0.5em},
% ylabel style={name=ylabel, yshift=-0.1em},
% %%%set up grid
% xmax = 100,
% grid = major,
% xtick={},
% ytick={0,0.2,...,1},
% grid style={dashed, gray!30},
% legend style={font=\footnotesize},
% legend pos=south east,
% ]
% \addplot+[mark options={mark size=0.5pt},line width=0.8pt] 
%   table[x=x, y=y, col sep=comma] {figure/conndur_cn_1.csv};
%   \addplot+[mark options={mark size=0.5pt},line width=0.8pt] 
%   table[x=x, y=y, col sep=comma] {figure/conndur_cn_1.1.csv};
% \legend{IP conn = 1, IP conn $>$ 1}
% \end{axis}
% \end{tikzpicture}
}
\caption{CDF of durations of connections. A connection with ``0'' duration means 
its duration is less than one second. X axis is truncated at 100.}\label{fig:dur_per_ip}
\end{figure}

\bheading{Connection duration.}
We found that a large number of incoming connections were short-lived
connections; we also discovered peers performing crawling and probing for
Bitcoin network by establishing connections with very small duration.

We first examined the duration of completed connections for all IPs; the
median was only 1.3 seconds; 93.9\% of connections were less than 60
seconds; and 34.6\% of the connections had duration less than 1 second.
Only about 0.5\% of completed connections lasted for more than 1 hour
and we found a large fraction of them came from a service called
BitcoinRussia. 
We compared durations for different types of connections.  As shown in
\figref{fig:dur_per_ip}, connections from unreachable IPs have
higher percentage of short-live connections, e.g., the median connection
duration from such IPs is 0.5 seconds and more than 70\% of the
connection durations are less than 1 second.

We examined connections with very small duration: we call a
connection \emph{ephemeral} if its duration is less than 0.5 second. 
A large portion (48.3\%) of connections from unreachable IPs are
ephemeral.  These connections account for 59.4\% of all ephemeral
connections~(760,948 in total).  Unavailable IPs contribute 34.7\% of
all ephemeral connections. 
% unreps: 451863; unavai 264313
We checked the client version strings in the ephemeral connections from
unreachable IPs and found 49\% of them are ``breadwallet'' or ``Bitcoin
Wallet'', which indicates these connections might issued by two mobile
Bitcoin applications: breadwallet and Bitcoin Wallet. We will show more
analysis results on these two applications in later section.  
%221619 out of 451863

For the ephemeral connections from unavailable IPs, 188,536 out of
264,313 (or 71\%) connections, are associated with two client version
strings: ``Snoopy'' or ``bitcoin-seeder''.  Note that Snoopy or
bitcoin-seeder are tools for performing crawling and probing for Bitcoin
network.  The client version strings suggest all of these clients might
be used for performing Bitcoin network crawling or probing.  More
surprisingly, 96\% of these suspicious probing connections from
unavailable IPs, which is 6.1\% of all completed connections, came from
4 IPs in two public clouds:   three IPs are in DigitalOcean and one IP
in AWS EC2. Actually, almost all the probing connections originated from
clouds were from only one unavailable IP in DigitalOcean, and some
Bitcoin developers already suspected that this IP was for conducting
unknown attacks against Bitcoin network. 
% 46.101.246.115, 138.197.144.194, 52.76.95.246, 104.236.95.174
% 180861/ 188,536

%%%%%%%%%%%%%%%%%%%%%%

\begin{table}[t]
\resizebox{\columnwidth}{!}{
\footnotesize
\centering
\begin{tabular}{|l|l|l|l|l|}
\hline
\textbf{Type} & \textbf{Service} & \textbf{Client version} & \textbf{\# conn} & \textbf{\%} \\ \hline
\hline
1 & BitcoinRussia & bt-russia.ru:0.0.1f  & 338,399 & 11.5\\ \hline
2 & Hetzner & bitnodes.21.co:0.1 & 222,594 & 7.6 \\ \hline
1 & Linnode & 8btc.com:1.0 & 194,285 & 6.5\\ \hline
1 & DigitalOcean & Snoopy:0.2.1 &  191,228 & 6.5 \\ \hline
0 & Google & bitcoin-seeder:0.01  &  53,843 & 1.8\\ \hline
1 & DigitalOcean & bitcoin-seeder:0.01  & 53,592 & 1.8 \\ \hline
2 & Amazon & bitcoin-seeder:0.01  & 52,491 & 1.8\\ \hline
2 & Amazon & bitcoin-seeder:0.01 & 52,312 & 1.8 \\ \hline
0 & Amazon & bitcoin-seeder:0.01  & 52,267 & 1.8\\ \hline
2 & Amazon & bitcoin-seeder:0.01  & 52,078 & 1.8\\ \hline
\end{tabular}
}
\caption{Top 10 IPs based on number of associated connections per IP. Type 0, 1, 2 
correspond to unreachable, unavailable, available IPs, respectively. Percentages 
are the fraction of the total number of completed connections.}
\label{top10ip}
\end{table}

\begin{figure}[t]
\centering
\resizebox{.8\columnwidth}{!}{\begin{tikzpicture}
\begin{axis}
[
y tick label style={
        /pgf/number format/.cd,
            fixed,
            fixed zerofill,
            precision=2,
        /tikz/.cd
    },
%%%basic
width = 0.8\linewidth, 
height = .6\linewidth,
ylabel = Percentage,
xmode = log,
xlabel = Number of connections,
xlabel style={name=xlabel, yshift=0.5em},
ylabel style={name=ylabel, yshift=-0.1em},
%%%set up grid
grid = major,
xtick={1, 10, 100, 1000, 10000, 100000},
ytick={0,0.2,...,1},
grid style={dashed, gray!30},
legend style={font=\scriptsize},
legend pos=south east,
]
\addplot+[mark options={mark size=0.5pt},line width=0.8pt, mark=*, color=colorbrewer1] 
  table[x=x, y=y, col sep=comma] {figure/connperip0.csv};
  \addplot+[mark options={mark size=0.5pt},line width=0.8pt, mark=square*, color=colorbrewer6] 
  table[x=x, y=y, col sep=comma] {figure/connperip1.csv};
  \addplot+[mark options={mark size=0.5pt},line width=0.8pt, mark=triangle*, color=colorbrewer5] 
  table[x=x, y=y, col sep=comma] {figure/connperip2.csv};
\legend{Unreachable, Reachable but Unavailable, Reachable and Available}
\end{axis}
\end{tikzpicture}}
\caption{CDF of the total number connections from a given IP.}\label{fig:conn_per_ip}
\end{figure}
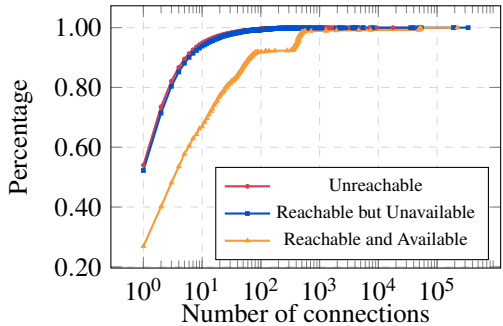

\bheading{Connection frequency.}
Next, we examine how many connections each unreachable peer made during
the measurement period. We found about 54\% of unreachable IPs only made
one connection, 95\% made less than 10, and 99\% made less than 50,
while unavailable IPs have a very similar long-tailed distribution~(see
\figref{fig:conn_per_ip}). In contrast, available IPs tend to make more
connections on average: only 37\% made one connection, while about 9\%
of them made more than 100 connections and 2\% made more than 500
connections. Some unreachable or unavailable IPs have made a large
number of connections, as shown in \tabref{top10ip}. As we can see from
\tabref{top10ip},  seven out of ten IPs belong to public clouds.

\begin{figure}[t]
\centering
\resizebox{.9\columnwidth}{!}{\begin{tikzpicture}
\begin{axis}
[
y tick label style={
        /pgf/number format/.cd,
            fixed,
            fixed zerofill,
            precision=2,
        /tikz/.cd
    },
%%%basic
width = 0.5\linewidth, 
height = .4\linewidth,
ylabel = Percentage,
xlabel = No. of IPs,
xlabel style={name=xlabel, yshift=0.5em},
ylabel style={name=ylabel, yshift=-0.1em},
%%%set up grid
grid = major,
xtick={},
ytick={0,0.2,...,1},
grid style={dashed, gray!30},
legend style={font=\footnotesize},
legend pos=south east,
]
\addplot+[mark options={mark size=1pt},line width=0.8pt] 
  table[x=x, y=y, col sep=comma] {figure/ip_per_tx.csv};
%\legend{Unresponisve}
\end{axis}
\end{tikzpicture}

\begin{tikzpicture}
\begin{axis}
[
y tick label style={
        /pgf/number format/.cd,
            fixed,
            fixed zerofill,
            precision=2,
        /tikz/.cd
    },
%%%basic
width = 0.5\linewidth, 
height = .4\linewidth,
ylabel = Percentage,
xlabel = No. of countries,
xlabel style={name=xlabel, yshift=0.5em},
ylabel style={name=ylabel, yshift=-0.1em},
%%%set up grid
grid = major,
xtick={1, 10, 20, 30, 36},
ytick={0,0.2,...,1},
grid style={dashed, gray!30},
legend style={font=\footnotesize},
legend pos=south east,
]
\addplot+[mark options={mark size=1pt},line width=0.8pt] 
  table[x=x, y=y, col sep=comma] {figure/tx_ip_reg.csv};
%\legend{Unresponisve}
\end{axis}
\end{tikzpicture}}
\caption{CDF of number of IPs \textbf{(left)} and number of countries \textbf{(right)} 
that involve in the propagation of  a given transactions.}\label{fig:ip_per_tx}
\end{figure}
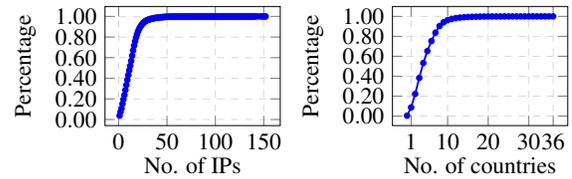

%amazon asn: 16509, 14061, 63949, 38895, 15169, 14618
\bheading{Transaction characteristics.} As shown in
\figref{fig:ip_per_tx}, a propagation could be relayed by as many as 211
IPs. The IPs that are involved with a given propagation could be
distributed across ASes and countries. For example, 1\% of propagations
are associated with IPs from more than 14 countries. We further found
propagations could be relayed by different types of peers: only 5.3\% of
the propagations are relayed by IPs of the same type, while 75.4\%  are
relayed by a mix of unreachable, unavailable, and available IPs.  

Another observation is that propagations are usually sent over long-live connections, e.g, 
99.9\% of the propagations were sent over the connections the durations 
of which were longer than 100 seconds. We grouped propagations based 
on source IP types and found the observations still hold true for different groups 
of propagations. Only 64 transactions were sent over ephemeral connections. 

\bheading{Characterizing different types of clients} In this section,
we focus on different types of peers: (1) \emph{mobile} peers, which are
the peers whose version strings contain ``breadwallet'' or ``Bitcoin
Wallet'', (2) \emph{probe} peers, which are whose version strings
contain ``Snoopy'' or ``bitcoin-seeder'', and (3) \emph{tor} peers,
which are the peers that also serve as Tor exit nodes or relay nodes.
They represent different types of clients. 

As shown in \tabref{tab:peer_stat_by_type}, mobile peers are associated
with 131,610 of unreachable IPs, which is 80\% of unreachable IPs.
These IPs contributed to about 20\% of all connections, most of which
were short-lived connections (less than 0.5 seconds).  We saw only 1.2\%
of propagations sent from these IPs; however, these propagations
correspond to 61\% of unique transaction IDs.  On the average, five
transactions were sent over a connection.  One possible explanation for
this phenomena is that these peers were configured not to forwarding
transactions.  

We observed a small number of IPs that we also identified as either Tor
exit or Tor relay nodes. Similar to mobile peers, they are involved in
less than 1\% of propagations but have sent over 41\% of transactions.
Compared to other types of peers, such tor peers have relatively longer
connection durations: the median duration of tor peers, probe peers,
and mobile peers are 2.1, 1.2, and 0.4, respectively 

No probe peers from available IPs have sent any transactions but
generated a large number of connections.

\begin{table*}[]
\centering
\resizebox{2\columnwidth}{!}{
\tabcolsep=0.03cm
\begin{tabular}{|l|l|l|l|l|l|l|l|l|l|}
\hline
 & \multicolumn{3}{c|}{\textbf{Mobile}} & \multicolumn{3}{c|}{\textbf{Probe}} & \multicolumn{3}{c|}{\textbf{Tor}} \\ \hline
 \hline
 & Unreachable & Unavailable & Available & Unreachable & Unavailable & Available & Unreachable & Unavailable & Available \\ \hline
 \#IP & 131,610~(69.6) & 16,862~(8.9) & 210~(0.11) & 12~(0.01) & 38~(0.02) & 490~(0.26) & 171~(0.09) & 295~(0.16) & 14~(0.01) \\ \hline
 \#Conn & 506,921~(17) & 67,356~(2.3) & 2,165~(0.07) & 97,627~(3.3) & 323,885~(11.0) & 567,069~(19.3) & 2,069~(0.07) & 5,819~(0.2) & 137 \\ \hline
 \#EphConn & 314,394~(10.7) & 39,724~(1.4) & 1,422~(0.05) & 195~(0.01) & 188,536~(6.4) & 1,011~(0.03) & 66 & 300~(0.01) & 71 \\ \hline
 % Med. Dur & \multicolumn{3}{l|}{0.4} & \multicolumn{3}{l|}{1.2} & \multicolumn{3}{l|}{2.1} \\ \hline
 % Med. \#Conn/IP & \multicolumn{3}{l|}{1} & \multicolumn{3}{l|}{2} & \multicolumn{3}{l|}{6} \\ \hline
 \#Prop & 1,008,567~(1.2) & 497,848~(0.6) & 386,557~(0.5) & 41,009~(0.05) & 154,393~(0.18) & 0 & 219,261~(0.26) & 533,865~(0.64) & 46,753~(0.06) \\ \hline
 \#TX & 1,518,206~(61.0) & 626,736~(25.2) & 679,264~(27.3) & 41,283~(1.7) & 168,589~(6.7) & 0 & 255,699~(10.3) & 717,077~(28.9) & 61,112~(2.5) \\ \hline
\end{tabular}
}
\caption{A breakdown of IP, connections and transactions for different types of peers.
``IP'', ``Conn'', ``EphCOnn'', ``Prop'', and ``TX''  stand for connection, 
unique IP addresses, connections, ephemeral connections, propagations, and
unique transaction ID respectively. Number 
in parenthesis is the percentage of the total number of all IPs/connections/propagations 
/unique transactions IDs, depending on statistics type.}
\label{tab:peer_stat_by_type}
\end{table*}

\textit{Limitation and discussion.}\label{sec:stats:limit}
We didn't verify if a transaction is valid or invalid. Doing
so requires a parser for Bitcoin blockchain information, and we are
working on such parser as part of bcclient analysis engine.

\section{On further enriching\\ the dataset}
\label{sec:timing}
In the previous section, we observed that a single transaction can be forwarded
by as many as 151 different IP addresses (excluding IPs of publicly known full
Bitcoin nodes) from a wide range of countries which makes extracting
per-country transaction origin statistics problematic: it would require one to
distinguish between the peer that actually created the transaction
(\textit{originator}) from the rest (i.e. peers that merely relayed it, we call
such peers \textit{relays}). This problem is well-suited for timing analysis
used in~\cite{neudecker2016timing} for inferring Bitcoin network topology of
publicly reachable peers. We are working on implementing the timing analysis
approach for unreachable clients as a part of the analysis engine of bcclient. As a
preliminary step, we want to understand how efficient this method would be for
unreachable peers, and design experiments to evaluate it. We find this
method to be quite efficient and reliable.

\begin{figure}[t]
\centering
\includegraphics[scale=0.4]{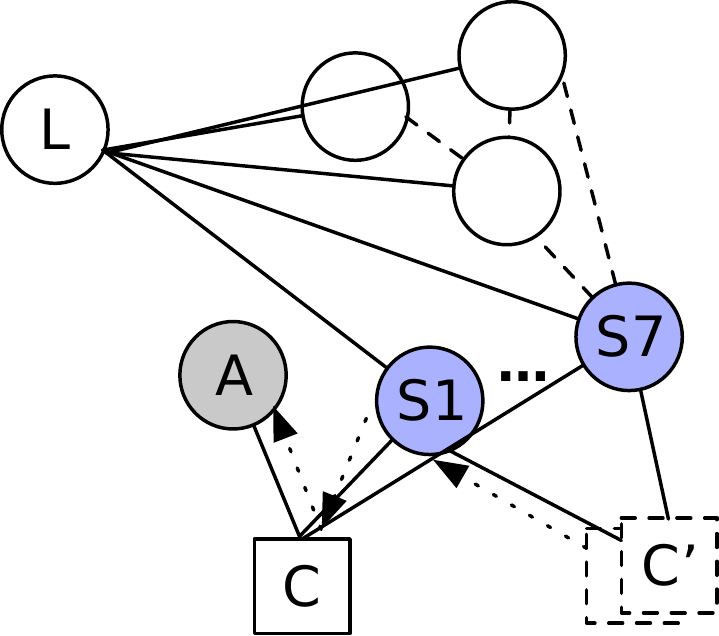}
  \caption{Distinguishing transaction true originator}
\label{fig:timingattack}
\end{figure}

To better explain the problem, consider a simple example in
\figref{fig:timingattack} with two peers $C$ and $C'$, that connect to
the Bitcoin network as regular unreachable clients; $C$ is connected
to one of our \emph{monitoring} nodes $A$ 
(which is a public Bitcoin peer but doesn't connect to any other peers). 
Assume $C$ and $C'$ generate transactions
$tx$ and $tx'$ both of which will be delivered to $A$ through $C$.
Node $A$ wants to distinguish transaction $tx$ (for which $C$ is true
originator) from $tx'$ (for which $C$ is a relay).  To do this we
introduce additional \textit{listener} node $L$ that is connected and listens to the
publicly known set of Bitcoin servers and use the following
criteria to distinguish between these two cases: \textit{If node $A$
receives $tx$ before $L$, we conclude that $C$ is the
originator for $tx$, and a relay otherwise}. 

\begin{figure}[t]
\centering
\includegraphics[scale=0.55]{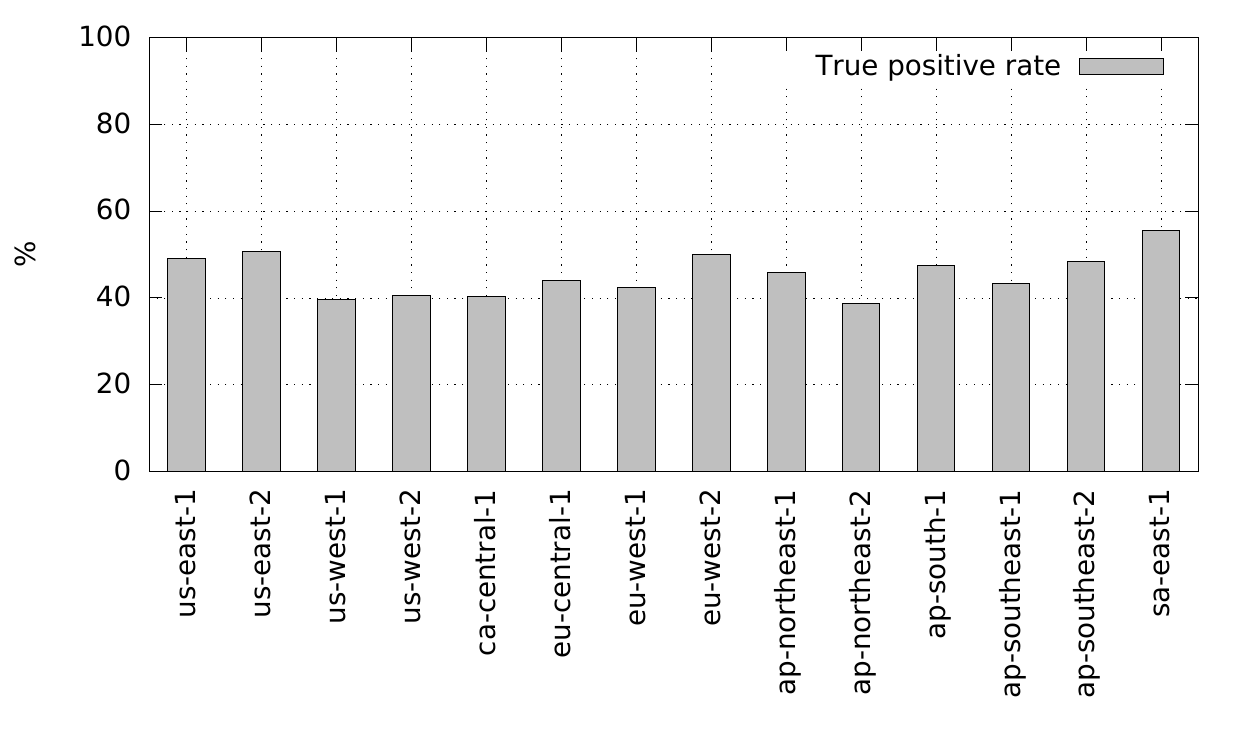}
\caption{True positive rate for different regions.}\label{fig:deanon-tp}
\end{figure}

\bheading{Experiments in the Testnet.}
We first evaluate true-positive rate of this method i.e. the chance of
successfully identifying $C$ as the originator of $tx$, by carrying out
experiments in the Bitcoin Testnet. For all the experiments, $C$ was running
Bitcoin core software version v0.14.1. In total we conducted 21 different
experiments in which we sent 8400 transactions from node $C$.  We moved nodes
$A$ and $C$ across 14 different geographical regions (see
Section~\ref{sec:stats}) and used transactions of different sizes~(i.e., the
number of receiver addresses in the transaction, chosen from 5, 10, 20, 50, and
100) in order to estimate how these factors affect the efficacy of our timing
analysis. We found that the true-positive rates varied from experiment to
experiment (from 20\% to 60\%), but on the average was consistent across
different regions and different transaction sizes. The resulting true positives
rates for each of the 14 geographical regions are shown in
Figure~\ref{fig:deanon-tp} with the average 45.43\% over all regions.

To estimate the false-positive rate, we set up client $C$ and
monitoring node $A$ in the same availability zone in \texttt{us-east-1} region.  We
count the number of transactions forwarded by node $C$ that appeared at $A$
before $L$. In order to estimate how number of concurrent connections affects
the false-positive rate, we make listener node $L$ establish different number
concurrent connections~(1, 2, 5, 10, 15, 20) to every peer in the Bitcoin network.  
For each setting, we record 5,000 transactions relayed by
\clnode, and examine how many of them are false positives.  We only discovered
a total of 12 false positives out of 30,000 transactions~(which amounts for
just 0.04\%). 

To better understand this result we further look at the transaction latency
(i.e., the time between $C$ creates a transaction and node $A$ receives it)
under different transaction sizes and region combinations, and find that in the
majority of cases client delay (see Section~\ref{sec:background}) dominates
network latency. For instance, even though $C$ and $A$ are in the same EC2 zone, 
the longest transaction latency can be as long as 8.5 seconds. 

\bheading{Experiments in the Mainnet.}
We now estimate true-positive and false-positive rates in the Bitcoin Mainnet.
We first place client node $C$ monitoring node $A$ in the same region
(\texttt{us-east-1}) and carry out two experiments in which we vary client's
entry nodes; we send 20 transactions from client $C$ in each of the
experiments. The resulting true positive rate is 65\% and 40\%.  Next we
place client $C$ and node $A$ in different regions (\texttt{us-east-1}
and ap-northeast-1) and make client $C$ send 24 transactions. The true
positive rate for this case is 41.6\%. These true-positive rates are
very close to the success rate we obtained earlier in the Testnet.  We
finally estimate the false-positive rate.  Out
of 25000 transactions only 20 arrived first at node $A$, which amounts to
0.08\% only. This proves that our method is a very reliable way to identify 
transaction origins.

\bheading{Discussion.}
Kaminsky was the first who described this general idea in his BlackHat
presentation~\cite{dakami} in the context of client deanonymization, and Koshy
et al.~\cite{koshyanalysis} was among the first to try to practically evaluate
it.  While he was able to deanonymize a number transactions that exposed
anomalous behavior (e.g.  transactions relayed only once or transaction that
were relayed  multiple times by the same IP), he concludes that  assigning a
transaction's ownership to its first relayer is ineffective. Neudecker et al.~\cite{neudecker2016timing}
 evaluate this approach in the mainnet but only consider public reachable
 peers. Compared to their results, we actually achieve much better performance.
 We believe the discrepancy with our results is due to that we considered
 unreachable peers and levereaged the recent modifications to Bitcoin core
 software.

\section{Conclusion}
\label{sec:conclusion}
In this paper, we developed an experimental framework which we used to
collect and analyze various statistics on Bitcoin unreachable peers, a
part of the Bitcoin network that despite its size did not receive much
attention from the research community but which is crucial to enhance
the scientific understanding of the Bitcoin system.

In our study we deployed monitoring nodes spread over the globe and
collected more than two million connections and transactions.  We find
several previously unreported and surprising results. First, we find
that Bitcoin unreachable peers appear to be centralized: a large number
of peer are hosted in few Autonomous Systems.  Second, the version
messages that we received from the clients as well their ISP's strongly
suggest that the overwhelming majority of Bitcoin users access Bitcoin
through a mobile application. Moreover, we found that those peers are
associated to two mobile Bitcoin applications, which means that security
of a large number of transaction is in hands of just two companies.
Third, we discovered a large fraction of connections came from public
clouds; these might be used by researcher and other interested
parties to monitor and crawl the network.
We further experimentally evaluated a timing analysis-based technique 
for triage the first relay that forwards a given transaction and our results 
suggest the technique can achieve near zero false-positive rates against 
unreachable Bitcoin peers. 

We believe that our work sheds more light into a more hidden part of the
Bitcoin network and is a valuable step towards better understanding the
Bitcoin ecosystem as a whole.

%%%%%%%%%%%%%%%%%%%%%%%%%%%%%%%%%%%%%%%%%%%%%%%%%%%%%%%%%%%%%%%%%%%%%%%%%%%%%%%%

\balance
\bibliographystyle{abbrv}
%\scriptsize
%\vspace{-0.2cm}
%\vspace{0.4cm}
\bibliography{bitcoin}
\end{document}